\documentclass[oldversion]{aa}

\usepackage{graphicx,t1enc}

\begin{document}
{

\title{A multicolor near-infrared study of the dwarf nova IP Peg}

\author{T. Ribeiro\inst{1},
  R. Baptista\inst{1}$^{,}$\inst{2},
  E. T. Harlaftis\inst{3}\thanks{in memoriam},
  V.S. Dhillon\inst{4}
  R.G.M. Rutten\inst{5}, 
  }

\institute{Departamento de Física , Universidade Federal de Santa Catarina, Campus Trindade, 88040-900, Florian\'opolis, SC, Brazil\\
  \email{tiago@astro.ufsc.br}
  \and
  Soar Telescope, Colina El Pino s/n, 
  Casilla 603, La Serena, Chile
  \and
  Institute of Space Applications and Remote Sensing, National Observatory of 
Athens, PO Box 20048, Athens 118 10, Greece
\and
Dept. of Physics and Astronomy, University of Sheffield, Sheffield S3 7RH, UK
  \and
Isaac Newton Group of Telescopes, Apartado de Correos 321, E-38700, Santa
Cruz de La Palma, Spain
}

\date{Received ???; accepted ???}

\abstract
{\normalsize
We report the analysis of $JHK_{s}$ light curves of the eclipsing dwarf nova 
IP Peg in quiescence. The light curves are dominated by the
ellipsoidal variation of the mass-donor star, with additional 
contributions from the accretion disc and anisotropic emission from the bright 
spot. A secondary eclipse is visible in the $J$ and $H$ light curves, with 
$2\%$ and $3\%$ of the flux disappearing at minimum light, respectively.  
We modeled the observed ellipsoidal variation of the secondary star (including
possible illumination effects on its inner face) to find a mass ratio of
$q = 0.42$ and an inclination of $i = 84.5^{o} $, consistent in the three bands
within the uncertainties. Illumination effects are negligible. The secondary is
responsible for $83\%$, $84\%$ and $88\%$ of the flux in $J$, $H$ and $K_{s}$, 
respectively. We fitted a black body spectrum to the $JHK_{s}$ fluxes of the 
secondary star to find a temperature of $T_{bb} = 3100\pm500\;K$ and a distance of 
$d=115\pm30$ pc to the system. We subtracted the contribution of the secondary star 
and applied 3-D eclipse mapping techniques to the resulting light curves to 
map the surface brightness of a 
disc with half-opening angle $\alpha$ and a circular rim at the radius of the 
bright spot. The eclipse maps show enhanced emission along the stream 
trajectory ahead of the bright spot position, providing evidence of gas stream 
overflow. The inferred radial brightness-temperature distribution in the disc
is flat for $R < 0.3R_{L1}$ with temperatures $\simeq 3500K$ and colors
consistent with those of cool opaque radiators.

}


\titlerunning{A multicolor near-infrared study of the dwarf nova IP Peg}

\authorrunning{Ribeiro et al.}

\maketitle

\section{Introduction}
\label{sec:int}
{\normalsize

The shape of the secondary star in
cataclysmic variables (CVs) and low-mass x-ray binaries (LMXBs) is defined by 
its Roche equipotential surface. Modeling the ellipsoidal variation produced in the light curves 
of these binaries by the changing aspect of the distorted secondary star 
with binary phase provides constraints on the orbital parameters 
(e.g., Warner 1995) and yields an estimate of the contribution of the 
secondary star to the total light. 

Unlike visible and ultraviolet light curves, which are dominated by
emission from the disc and bright spot, the near-infrared (NIR) light
curves of CVs and LMXBs have significant contributions from the
cool secondary star.
The determination of the secondary star flux in a set of wavelengths allows a 
direct estimate of its spectral type and also opens up the possibility of 
inferring the distance to the binary by means of photometric 
parallax. Furthermore, knowing the contribution of the secondary star to the 
light curve allows one to isolate the light from the accretion disc and to 
apply eclipse mapping techniques (e.g., Baptista \& Steiner 1993) to 
investigate the disc structure in the infrared.

IP Peg is a relatively bright ($V \simeq 14.5$ mag), long period 
($P_{orb} = 3.8$ hr) dwarf nova showing recurrent outbursts every few months in
which the system increases its brightness by $\simeq$ 2 mag in the visible. The
binary is seen almost edge on (inclination $i>80^{o}$) allowing for 
eclipses of both the white dwarf/accretion disc and the secondary star,
as well as of the bright spot (hereafter BS). Optical and ultraviolet (UV) 
light curves of IP Peg are dominated by anisotropic emission from the prominent 
BS, making it hard
to constrain the system parameters (e.g., Wood \& Crawford 1986) and, 
therefore, difficult to apply eclipse mapping techniques to 
investigate its accretion disc structure. Szkody \& Mateo (1986) found evidence
of ellipsoidal variations in $JHK_{s}$ light curves of IP Peg. Froning et 
al. (1999) [hereafter F99] found a pronounced double-hump modulation after 
subtracting the ellipsoidal variation of the secondary star from their $H$ band
light curve. Their eclipse mapping analysis indicates that the accretion disc 
is cool with a flat radial temperature distribution ($T_{bb} \simeq 3000\;K$) 
in the $H$ band.

This paper reports a multicolor study of IP Peg in the $JHK_{s}$ bands. 
Section \ref{sec:bialc} presents the observations and data reduction. 
Section \ref{sec:fac} describes the procedure used to
fit the ellipsoidal variation and the application of 3-D eclipse mapping
techniques to the light curves after we subtract the contribution of
the secondary star. The results are discussed in Section \ref{sec:disc} 
and summarized in Section \ref{sec:concl}.
 
}

\section{Observations and data reduction}
\label{sec:bialc}
{
IP Peg was observed with WHIRCAM (Hughes et al. 1996) at the $4.2$m William 
Herschel Telescope in La Palma on 1996 October 26-29, while the system was in 
quiescence. The observations comprise 3 orbital cycles in the $H$ band, and a 
bit less than 1 orbital cycle each in the $J$ and $K_{s}$ bands. All runs were 
performed in good weather with bright moon, but no clouds. The seeing ranged 
from 1.0'' to 1.6''. The exposure times were of 1-3s ($H$) and of 3s ($J$ and 
$K_{s}$). The observations are summarized in Table \ref{log_tab}.
 
We used a dithering procedure to estimate the contribution of the sky 
background, nodding the telescope in a five-position square pattern
(center plus four corners). 
The sky level was obtained from the median of the five images of each set and was
then subtracted from each image of the set. A series of dark images, with the 
same exposure time as the science frames, were obtained during the run at 
intervals of $\sim 1-2$ hrs. These images were used to remove the dark current 
of the chip. Correction of flat-field effects was also performed.

Data reduction was performed using APPHOT/IRAF\footnote{IRAF is distributed by 
the National Optical Astronomy Observatories,
which are operated by the Association of Universities for Research
in Astronomy, Inc., under cooperative agreement with the National
Science Foundation.} aperture photometry routines. 
The frames of each dithering sequence were aligned with an interactive 
procedure based on the IRAF 'register' task and combined to increase the 
signal-to-noise ratio (S/N) of the measured stars. The non-linearity of the 
detector was corrected using the 'irlincor' routine in the CTIO package.  
Fluxes were then extracted for the variable and for a comparison star 
$7.4 \arcsec$ North and $7.5 \arcsec$ East of the variable.

Differential light curves (target star flux divided by comparison star flux)
were computed. The light curves were flux calibrated using the 2MASS $JHK_s$ zero 
point constants (Skrutskie et al. 2006) and the absolute magnitudes for the comparison 
star ($J = 9.82$, $H = 9.56$ and $K_s = 9.48$). No color term corrections were applied. 
The data were phase-folded according to a modified version of the linear ephemeris of Wolf et
al. (1993), 
\begin{equation}
  T_{mid}(HJD) = 2445615.4156(4)+0.15820616(4) \cdot E,
\label{efem}
\end{equation}
were $T_{mid}$ is the time of inferior conjunction of the secondary star. The 
resulting light curves can be seen in Fig. \ref{cur_luz}. The $H$ band data 
were combined to improve the $S/N$ of the light curve. The $J$ and $K_{s}$ band
data do not cover the full orbit of the binary.  

\begin{table}
\begin{center}
\caption{\label{log_tab} Journal of the observations.}
\begin{tabular}{cccccc}
\hline\hline
Date & band &  start  & No. of & Eclipse$^1$& phase	  \\
1996 &      &   (UT)  & frames &  cycle     & range	  \\ [1ex] \hline
10/26& $H$  &  19:00  &  1362  &  30138     & -0.86,+0.50 \\
     &      &         &        &  30139     & -0.50,+0.20 \\
10/27& $H$  &  20:00  &   760  &  30144     & -0.28,+0.56 \\
10/27& $K_{s}$  &  23:16  &   590  &  30145     & -0.42,+0.24 \\
10/28& $J$  &  19:15  &  1212  &  30150     & -0.15,+0.50 \\
     &      &         &        &  30151     & -0.50,+0.34 \\ [1ex]
\hline   
\multicolumn{6}{l}{ $^1$ -  with respect to the ephemeris of eq.\ref{efem}}\\
\end{tabular}
\end{center}
\end{table}

}

\section{Data analysis}
\label{sec:fac}
{

\subsection{The white dwarf eclipse width $\Delta\phi$.}
\label{sec:wdew}
{
The optical study of Wood \& Crawford (1986) leads to a white dwarf 
eclipse width in the range $\Delta\phi = 0.086-0.092$. The uncertainty
in estimating $\Delta\phi$ arises from the fact that the white dwarf eclipse 
ingress in IP Peg is veiled by the much more pronounced ingress of the 
BS eclipse and by the large amplitude flickering seen in optical light 
curves prior to BS ingress.  

In order to refine the value of $\Delta\phi$, we employed an 
iterative procedure, assuming binary parameters derived from the fit 
of the ellipsoidal variation (Sect. \ref{sec:ellip}) and applying the 
corresponding phase offset $\phi_0$ needed to make the observed white dwarf 
mid-egress feature coincident with phase $\Delta\phi/2$. The procedure searches
for the pair of ($\Delta\phi$, $\phi_0$) values that yields the best $\chi^2$ 
for the model light curve. This iterative 
process converges to a solution with $\Delta\phi = 0.0918$ and
$\phi_0 = +0.01$, consistent with the upper limit of Wood \& Crawford (1986).
We use the $J$ band light curve in this process because it shows clearly 
the white dwarf egress, unlike the $H$ band light curve. The $K_{s}$ band light
curve was not used for this procedure because of its lower S/N and reduced 
phase coverage. Because the white dwarf egress feature is present in both 
J band eclipses, we may safely exclude the possibility that the observed 
feature is due to a flicker/flare from the disc. The light curves shown in Fig. 
\ref{cur_luz} were corrected by the derived value of $\phi_0$ in order to make 
mid-eclipse coincident with phase zero.

\begin{figure}	
\begin{center}
\scalebox{1}[1.00]{%
\includegraphics[bb=0.1cm 3cm 20cm 24.5cm,scale=0.36,angle=270]{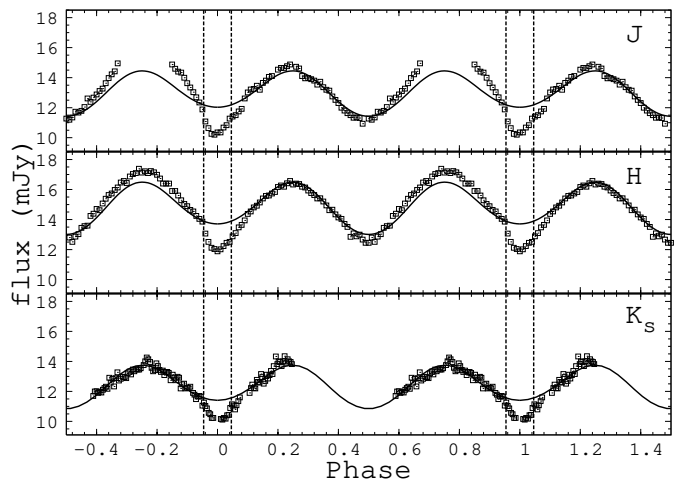} }
\caption{Light curves of IP Peg with modeled ellipsoidal variation added to 
the inferred (constant) disc contribution to fit the data.
Vertical dashed lines mark the ingress/egress phases of the white dwarf for an 
eclipse width of $\Delta\phi = 0.0918$ cycles.}
\label{cur_luz}
\end{center}
\end{figure}

}
\subsection{Fitting the ellipsoidal variation.}
\label{sec:ellip}
{
The ellipsoidal variation is modeled with the aid of a light-curve synthesis 
program, which computes the flux emitted by a Roche-lobe filling star as its 
aspect changes with binary phase. The surface of the Roche lobe filling 
star is divided into a large number of tiles. The flux emitted by each tile
is modified to account for gravity- and limb-darkening effects. We used the 
gravity-darkening coefficient of Sarna (1989), $\beta = 0.05$,
and adopted the non-linear square-root limb-darkening law of Diaz-Cordovés \& 
Giménez (1992),
\begin{equation}
  I = I_{0}(1-a(1-cos\gamma)-b(1-\sqrt{cos\gamma})),
\end{equation}
where $\gamma$ is the angle between the line of sight and the normal 
to the surface of the tile, and $I_{0}$ is the intensity emitted normal to the
tile surface. We assumed $a_{J} = -0.465$, $a_{H} = -0.454$, $a_{K} = -0.448$ 
and $b_{J} = 1.199$, $b_{H} = 1.173$, $b_{K} = 1.066$ (Claret 1998).
 
In order to account for both limb and gravity darkening we need to adopt values
for the temperature $T_{pole}$ and gravity $g_{pole}$ at the pole of the star.
Using a procedure similar to that described by F99, we tested a 
range of physically plausible values varying $T_{pole}$ from $2800$ to $3200 K$
and log$ g $ from $3.5$ to $5.0$. The results were rather insensitive to the 
choice of parameters, except for the $K_{s}$ band in which the program only 
converges for values of 
$T_{pole} = 3000 - 3200K$ and log$g = 4.5$. We then decided to adopt  
$T_{pole} = 3000 K$, log $g = 4.5$ and solar metallicity, since these values
are in better agreement with those in the literature 
(Szkody \& Mateo,1986; Leggett, 1992; F99).    

The parameters fitted by the ellipsoidal variation code are: mass ratio ($q$), 
binary inclination ($i$), the peak intensity of a spot at the inner 
hemisphere of the secondary star relative to the mean intensity at the star's
surface ($I_s/I_0$), the orientation angle of the spot with 
respect to the line joining both stars ($p_s$), a phase-independent
additive flux level , and the flux of the secondary star at phase zero ($F_s$).
The program minimizes the $\chi^{2}$ between the observed light curve and 
the ellipsoidal variation model for a given set of parameters with an 
{\it amoeba} 
minimization scheme (Press et al. 1986). Alternatively, it is possible to fit a
phase offset that minimizes the $\chi^{2}$ of each set of parameters 
(see Sec. \ref{sec:wdew}).

The NIR light curves of IP Peg show contributions from other sources apart from
the ellipsoidal variation of the secondary star. As a first step, 
we removed only the phases covering the
primary and secondary eclipses from the light curve before attempting to fit
the ellipsoidal variation. This leads to hard-to-converge, unrealistic 
solutions which underestimate the hump at phase $+0.75$ and overestimate the 
hump at phase $+0.25$ in the $J$ and $H$ bands.
The difference in brightness is larger in the $J$ band -- where the asymmetry 
in the eclipse shape caused by the BS is larger -- and disappears 
in the $K_{s}$ band, which shows almost no evidence of the BS in the eclipse 
shape. This indicates
that anisotropic emission from the BS contributes significantly
to the hump centered at phase $+0.75$. We therefore decided to omit the phase 
range $[+0.6;+0.9]$, as well as the primary and secondary eclipses, from the fitting 
procedure. With the new restriction, it was possible to fit a model light 
curve to the data of all bands.

\begin{table}
\begin{center}
\caption{\label{tab_result} Modeled IP Peg parameters}
\begin{tabular}{ccccc}
\hline\hline
          & $J$           & $H$           & $K_{s}$ \\[1ex] \hline
$i$       & $84^o\pm5^o$  & $85^o\pm3^o$  & $87^o\pm2^o$ \\
$q$       & $0.43\pm0.10$ & $0.42\pm0.20$ & $0.44\pm0.03$ \\
$F_s$ (mJy)& $9.9\pm2.0$& $11.5\pm2.5$ & $10.1\pm1.6$ \\
$F_s/F_T$ & $(82\pm5)\%$  & $(84\pm5)\%$  & $(88\pm4)\%$ \\
$I_s/I_0$ & $0.03\pm0.02$ & $0.04\pm0.03$ & $\leq 0.01$\\
$p_s$     & $(3\pm2)^o$   & $(2\pm1)^o$   &     --    \\[1ex]
\hline   
\multicolumn{4}{l}{$F_s$ is the flux of the secondary and $F_T$ is the total flux}\\
\end{tabular}
\end{center}
\end{table}

Table \ref{tab_result} lists the results of the fitting procedure. The model 
light curves are shown as solid lines in Fig. \ref{cur_luz}. Our results 
indicate that the secondary star is responsible for $84\%$ and $88\%$ of the 
total brightness in the $H$ and $K_{s}$ bands, respectively, in 
agreement with the results of Szkody \& Mateo (1986) and F99. 
Littlefair et al. (2001) fitted an M4V type star 
to the $K_s$ band spectrum of IP Peg to find that the secondary star 
contributes $62\%$ of the total light at that wavelength.  However,
as pointed out by Harrison et al. (2005a, b), the depth of the
absorption lines from the secondary star in CVs are reduced with
respect to isolated stars of similar spectral type, perhaps by some
atmospheric effect. In this case, attempts to match the depth of the
absorption lines will systematically underestimate the contribution 
of the secondary star to the total light. This effect may account for 
the lower contribution inferred by Littlefair et al. (2001). Alternatively, the
different inferred contributions could be a consequence of changes is disc 
brightness with time, with a slightly brighter disc leading to a lower 
secondary star relative contribution at the epoch of the observations of 
Littlefair et al. (2001). 
It is hard to test this possibility because IP Peg is at the limit of detection
for amateur astronomer while in quiescence, showing a typical scatter of $0.2$ 
mags in its historical light curve. We analyzed the AAVSO historical light 
curve of IP Peg and found no discernible (larger than $0.2$ mag) difference in 
brightness state between the epochs of the observations of F99,
Littlefair et al (2001) and this work.

The fitted Gaussian spot on the inner face of the secondary star is roughly 
centered at the L1 point and gives a negligible contribution to the total flux 
in all bands within the uncertainties, indicating that irradiation effects are
not significant in the IR continuum.
   
An inclination versus mass ratio diagram for IP Peg is shown in Fig. 
\ref{ixq}, where the derived range of values of $q$ and $i$ for the three
bands are depicted. There is 
good agreement between the results from the three bands, with the dispersion of
the mass ratio values being much smaller than the formal errors of the 
determinations in the $J$ and $H$ bands. The best fit (lower 
$\chi^2$ value) is obtained for the H band data. Because this light curve 
is the average of data from three orbits, it has the most complete
phase coverage, higher S/N, and lower influence from flickering.
The $J$ and $K_s$ light curves include data from only one orbit with 
incomplete phase coverage, although the lack of phase coverage of
the J band data does not affect the fit of its ellipsoidal variation.
In spite of their smaller formal error, the results for the $K_{s}$ band are 
less reliable since only a small part of the light curve was used for the fit.
Taking into account the constraint derived from the inferred width of the 
white dwarf eclipse (Sect. \ref{sec:wdew}), our best solution is $i = 84.5^{o}$
and $q = 0.42$. This set of parameters is indicated by a filled circle in 
Fig. \ref{ixq}.

Figure \ref{ixq} also compares our results with those in the literature. The
binary parameters derived from the ellipsoidal variations in the $J$ and $H$ 
band data are in reasonably good agreement with those of Wood \& Crawford 
(1986), Beekman
et al. (2000) and Watson et al. (2003). The solution of Marsh (1988) relies on
inferring the white dwarf radial velocity from the wings of the emission lines
from the disc. This technique is prone to large uncertainties, particularly
when there is significant phase offset between the spectroscopic inferior
conjunction of the secondary star and the observed mid-eclipse time - as it is 
in the case in IP Peg.
}

\begin{figure}	
\begin{center}
\scalebox{1}[1.00]{%
\includegraphics[bb=1cm 5cm 20cm 24.5cm,scale=0.37,angle=270]{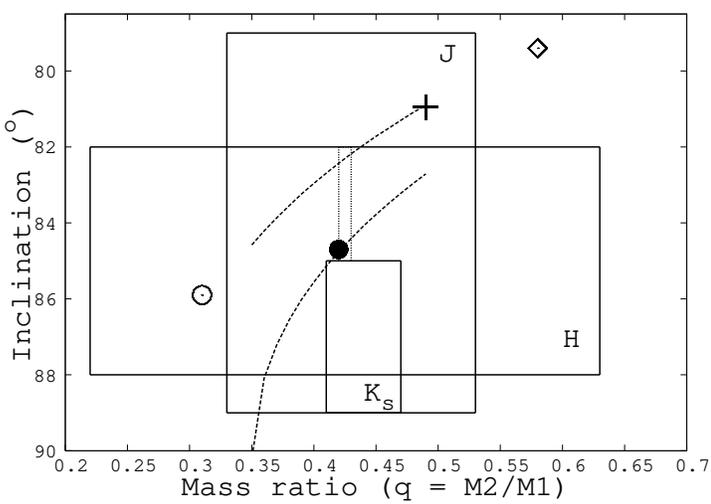} }
\caption{Inclination - mass ratio diagram. Dashed lines indicate the relations 
for $\Delta \phi = 0.0863$ (upper curve) and $\Delta\phi=0.0918$ (lower curve)
and bracket 
the range of (i,q) values of Wood \& Crawford (1986). Vertical dotted lines 
mark the range of values determined by Watson et al. (2003).
Three boxes indicate the $1-\sigma$ limits on the derived parameters from the 
$JHK_{s}$ band data. 
A cross, an open circle, an open diamond and a filled circle depict the 
formal solution of Wood \& Crawford (1986), Beekman et al (2000), Marsh (1988)
and this work, respectively. 
}
\label{ixq}
\end{center}
\end{figure}

Stellar atmosphere models (e.g. Kurucz 1979) and available stellar
atlases (e.g. Strecker et al. 1979) are not suitable for fitting IR fluxes of 
very cool stars, either because of a poor spectral resolution and sampling in 
the IR or because the range of temperatures of the grid/atlas does not include
stars cooler than $\simeq 3300K$. Therefore, we fitted black bodies to the 
extracted $JHK_s$ fluxes of the secondary star of IP Peg in order to estimate 
its temperature and distance. The best fit solution yields 
$T_{bb} = 3100\pm500K$ and a distance of $d_{bb} =(115\pm30) pc $, 
assuming $R_{2} = 0.4 R_{\odot}$ (Beekman et al. 2000). The results are
consistent with those of Szkody \& Mateo (1986).

}

\subsection{3-D eclipse mapping}
\label{sec:3dem}
{
After subtracting the ellipsoidal variation corresponding to the adopted binary
parameters from each light curve, eclipse mapping techniques (Baptista
\& Steiner 1993) were applied to the residual curves to derive maps of the 
surface brightness distribution of the IP Peg accretion disc in $JHK_{s}$. 

Our eclipse map is a 3D surface consisting of a 51 x 51 pixels disc grid with a
half-opening angle $\alpha$ 
(the angle between the midplane and the disc surface) covering the primary
Roche lobe (up to $R = R_{L1}$)
plus a circular rim of $101$ pixels at $R = R_{bs}$,
the inferred radius of the BS.
With this geometry there is no need to extract the orbital hump from the light 
curve (F99), since it can be accounted for by emission from the circular rim.
We also note that this geometry allows the disc brightness distribution to
extend beyond the radial position of the BS.
The radius of the BS was derived as follows. We measured the BS mid- 
ingress/egress phases ($\phi_{bi} = -0.027, \phi_{be} = +0.983$) in the $J$ 
band light 
curve after correcting for $\phi_0$ (Sect. \ref{sec:wdew}). For the adopted 
binary geometry (i,q), the pair of ($\phi_{bi}, \phi_{be}$) values maps into 
an x-y position in the orbital plane which consistently falls along the 
ballistic stream trajectory. 
The BS radius is taken as the radius of the circle that passes through this
position, $R_{bs} = 0.58R_{L1}$. The adopted binary geometry and the BS 
radius are depicted in Fig. \ref{sz}.  
For the eclipse mapping modeling, only 
the data in the phase range [-0.15;0.15] was analyzed.

We performed Monte Carlo simulations to estimate the uncertainties of the 
reconstructions (Rutten et al. 1992). For a given light curve a set of 100 
artificial light curves is
generated, in which the data points are independently and randomly
varied according to a Gaussian distribution with standard deviation
equal to the uncertainty at that point. The light curves are fitted
with the eclipse mapping code to produce a set of randomized eclipse
maps. These are combined to produce an average map and a map of the
residuals with respect to the average, which yields the statistical
uncertainty at each pixel. The uncertainties obtained with this
procedure are used to draw the contour maps of Fig. \ref{3dm}, and to estimate 
the uncertainties in the derived radial brightness-temperature distributions 
(Fig. \ref{bri_prof}) and the flux-ratio diagram (Fig. \ref{ccd}).

The disc half-opening angle $\alpha$ is a free parameter in the problem. The
entropy of the eclipse map is a useful tool in gauging the correct value of 
$\alpha$. Overestimating (underestimating) the disc half-opening angle 
introduces 
spurious structures in the disc side closest to (farther from) the L1 point. 
Because the entropy is a measurement
of the smoothness of the eclipse map, these structures are flagged with 
higher entropy. Therefore, one may estimate the value of $\alpha$ by performing
a set of reconstructions for a plausible range of $\alpha$ and selecting the
one with lowest entropy. 
Simulations (Borges, Baptista \& Catalan 2007) show that it is
indeed possible to use the entropy as a criterion to infer the value of
$\alpha$, but that the map of lowest entropy underestimates the correct
$\alpha$ by $1.5-2.0^{o}$, depending on the binary geometry $(i,q)$.
We confirmed this finding with careful simulations done with the specific
geometry of IP Peg, for which we find the offset to be $1.5^{o}$. 
In order to infer the value of $\alpha$ we used the 
$K_{s}$ band light curve, which has the most symmetric eclipse shape and 
the smallest contribution from the orbital hump. We tested a range of $\alpha$
values between $0^o - 8^o$ with increasing steps of $\Delta\alpha = 0.5^o$.
The map of lowest entropy and highest degree of symmetry is obtained for
$\alpha = 0.5^o$. According to the above simulations, we adopted $\alpha = 2^o$
for the eclipse mapping reconstructions. 
This is consistent with computations of vertical disc structure by 
Meyer \& Meyer-Hofmeister (1982), Smak (1992) and Huré \& Galliano (2000),
which predict $\alpha \simeq 1.5^{o} - 2.0^{o}$ for mass accretion rates 
\.{M} $= 10^{-11} - 10^{-10} M_{\odot}yr^{-1}$.

\begin{figure*}	
\begin{center}
\scalebox{1}[1.00]{%
\includegraphics[bb=1cm 5cm 19.5cm 24.5cm,scale=0.7,angle=270]{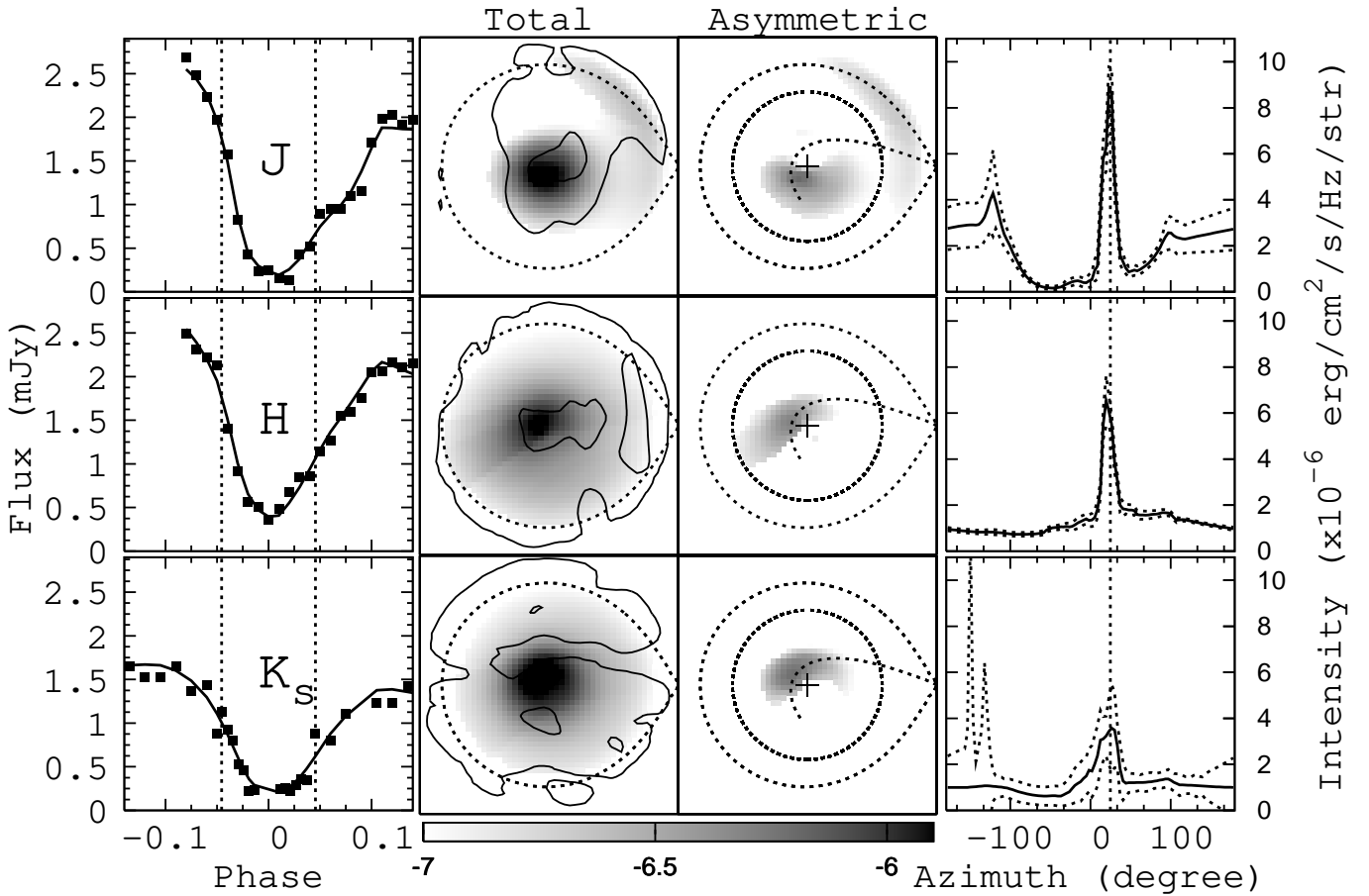} }
\caption{\textbf{Left:} The light curves after correction of ellipsoidal
variation (dots) and the best fit 3-D eclipse mapping model (solid curve).
Vertical dotted lines mark the ingress/egress phases of the white dwarf.
\textbf{Middle-left:} Corresponding disc brightness distributions on a 
logarithmic gray-scale. Dotted lines show the Roche lobe.
A solid contour line is overploted on each eclipse map to indicate the 
3-$\sigma$ confidence level on the intensities. The logarithmic intensity 
scale is indicated by the horizontal gray-scale bar (in $erg/cm^2/s/Hz/str$).
\textbf{Middle-right:} Asymmetric disc brightness distributions. Dotted lines
show the Roche lobe, the ballistic stream trajectory and a disc of radius 
$0.58R_{L1}$.
\textbf{Right:} Disc rim brightness distributions. The x-axis indicates the 
azimuth with respect to the line joining both stars. A vertical dotted line 
marks the azimuthal position of the bright spot. The dotted lines show the 
1-$\sigma$ limits on the intensities.}
\label{3dm}
\end{center}
\end{figure*} 

Light curves, corresponding eclipse maps and disc rim brightness distributions 
are shown in Fig. \ref{3dm}. For a better visualization of structures in the 
disc brightness distributions the asymmetric disc components are also shown.
A symmetric component is obtained by slicing the disc into a 
set of radial bins and fitting a smooth spline function to the mean of the
lower half of the intensities in each bin.  The
spline-fitted intensity in each annular section is taken as the symmetric
disc-emission component. This procedure preserves the baseline of the
radial distribution while removing all azimuthal structure. The
asymmetric disc component is then obtained by subtracting the symmetric
disc from the original eclipse map (Saito \& Baptista 2006). The asymmetric
disc component accounts for $30\%$, $20\%$ and $16\%$ of the total flux in 
$J$, $H$ and $K_s$, respectively.

The maps show enhanced emission along the ballistic stream trajectory close to
the white dwarf, providing evidence for gas stream overflow in IP Peg. The maximum of
the enhanced emission is wavelength-dependent, occurring farther downstream in 
$J$ than in $K_s$, possibly because of the progressively larger amount of 
gravitational energy released in collisions as the stream approaches the white 
dwarf. This tail of enhanced emission along the ballistic stream ahead of the 
BS position is reminiscent of that seen in the dwarf nova WZ Sge 
(Spruit \& Rutten 1998; Skidmore et al. 2000). Spruit \& Rutten (1998) point 
out that such a tail is to be expected as a consequence of the post-impact 
hydrodynamics of the stream.

A BS is expected to form at the intersection of the
ballistic stream with the disc outer edge. The ballistic stream hits 
the $0.58 R_{L1}$ disc rim at an azimuth $\theta=15.6^{o}$ with respect to 
the line joining both stars \footnote{The difference between the value 
of $\theta$ for the adopted q=0.42 geometry and for the alternative 
q=0.5 case (Wood \& Crawford 1986) is negligible.}. The $JHK_s$ disc rim 
distributions (Fig. \ref{3dm}) show a bright spot at an azimuth of $\theta 
\simeq 20^{o}$ with azimuthal extent (full width half maximum) 
$\Delta \theta \simeq 30 ^{o}$, in reasonable agreement with the 
predicted position of the BS.  The intensity of the spot decreases 
with increasing wavelength, in accordance with the corresponding 
reduction in the height of the orbital hump.  There is also a tendency 
for the centroid of the spot to move towards shorter azimuths with
increasing wavelengths.

It is interesting to compare the azimuthal position of the BS in 
the disc rim map with that of the orbital hump maximum. The azimuth 
of hump maximum is $\theta \simeq 74^{o}$ ($\phi_{max}\simeq -0.2$, 
Fig. \ref{cur_luz}), or $\simeq 54^{o}$ forward from the azimuth of the BS.  
Figure \ref{sz} shows a schematic diagram of IP Peg, where the offset between 
the azimuth of the BS and that of the maximum hump emission is clear. A similar
effect has been seen in other dwarf novae.  If BS 
emission is produced in a shock between disc and stream gas, the 
maximum emission will be normal to a direction between the disc and 
stream flows (Warner 1995, p.81).  For example, in OY Car the hump maximum 
is displaced $\simeq 19^{o}$ forward with respect to the azimuth of 
the BS (Wood et al 1989).  The offset in IP Peg is larger. Here, the 
azimuth of orbital hump maximum coincides with the normal to the gas 
stream at the BS position, suggesting that the impact shock lays along 
the stream trajectory with negligible influence from disc material.
We note that the phase of maximum of the orbital hump in the NIR-light curves
does not coincide with the one in the optical ($\phi = 0.9$, see Fig. 1 of Wood
\& Crawford 1986), as already noticed by F99.

\begin{figure}	
\begin{center}
\scalebox{1}[1.00]{%
\includegraphics[bb=3cm 5cm 20cm 24.5cm,scale=0.5,angle=270]{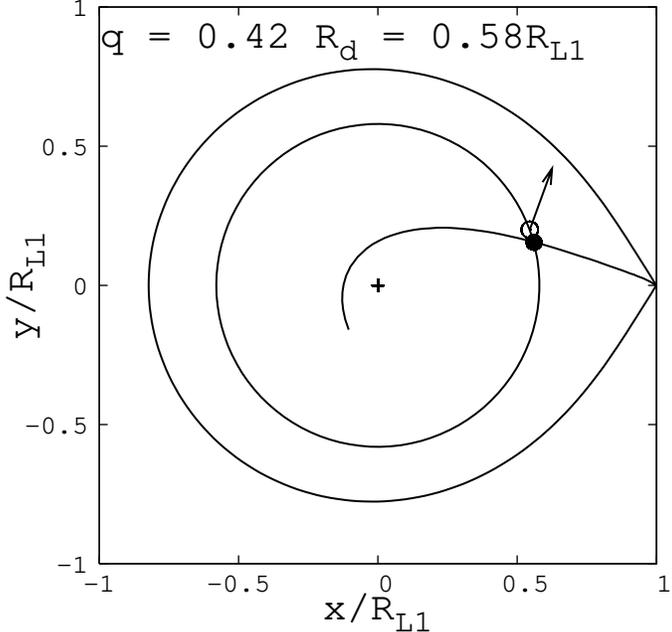} }
\caption{ The geometry of IP Peg. A filled circle marks the expected position 
of the BS; an open circle depicts the observed azimuth of maximum emission 
along the disc rim. An arrow indicates the direction of maximum emission of
the orbital hump, tilted forward $54^{o}$ from the azimuthal position of the 
bright spot. The large circle corresponds to a disc of radius $0.58R_{L1}$.}
\label{sz}
\end{center}
\end{figure}

Figure \ref{bri_prof} shows the $JHK_s$ disc radial brightness temperature 
distributions for an assumed distance of $d=115$ pc. Dotted lines show
the $T \propto R^{-3/4}$ law of opaque steady-state discs for mass
accretion rates of \.{M}$= 10^{-10}, 10^{-11}$ and $10^{-12}M_{\odot} yr^{-1}$,
assuming $M_1= 1.05 M_\odot$ (Beekman et al. 2000) and $R_1= 0.0124 R_\odot$ 
(derived using the white dwarf mass-radius relation of Nauenberg 1972).

The inner disc ($R < 0.3 R_{L1}$) shows a flat brightness temperature 
distribution reminiscent of those seen in quiescent dwarf novae (e.g.,
Wood et al. 1986, 1989), with inferred temperatures of $\simeq 3500$ K at 
$R=0.1R_{L1}$ for the $JHK_s$ bands within the uncertaities. 
The consistency of the inferred brightness temperatures in the three bands 
suggests that the inner disc is optically thick with emission close to 
blackbody.
The temperatures decrease in the outer disc 
regions ($R > 0.3 R_{L1}$) in reasonable agreement with the 
$T \propto R^{-3/4}$ law in the $J$ and $K_s$ bands and at a lower shallow gradient in the
$H$ band. There is marginal evidence\footnote{The observed difference in 
temperature is at the $1\sigma$ limit.} of higher brightness temperatures 
in the outer disc regions in 
the $H$ band with respect to the $J$ and $K_s$ bands. If real, such $H$ band
excess would indicate that the gas in the outer disc regions is opaque with a 
vertical temperature gradient (such that the $H^-$ free-free and bound-free 
opacity minimum at 1.6 microns leads to a relative increase in outcoming
flux in the $H$ band with respect to the $J$ and $K_s$ bands).
This is hard to reconcile with the results of Littlefair et al (2001), 
which suggest that the outer regions of the accretion disc in IP Peg are 
optically thin, unless the source of the observed mirror eclipses is an
optically thin chromosphere above the (optically thick) accretion disc.

A flux-ratio diagram of the accretion disc of IP Peg is shown in Fig.
\ref{ccd}. The colors for the symmetric component of 
the eclipse maps are plotted together with 
relationships for blackbody (BB), main-sequence stars (MS) and optically thin 
H\,I emission (HI). The BB and HI spectra were computed with the 
synphot/IRAF package. The BB and H\,I fluxes are extracted by convolving the respective 
spectrum with the response function of each infrared passband, and the flux 
ratios are computed. The MS colors were extracted from Bessell \& Brett 
(1988) and transformed to the 2MASS photometric system using the relations of
Carpenter (2001).
The uncertainties in the disc colors are 
quite large and increase towards the outer (and fainter) disc 
regions. The uncertainties in the colors are dominated by the lower S/N 
of the $K_s$ band light curve and eclipse map.
The colors of the symmetric inner disc (R<0.3 $R_{L1}$)  
are consistent with those of cool opaque radiators (with $T_{color} \simeq 
3000-5000\; K$) at the 1-$\sigma$ confidence level. The disc becomes redder 
(cooler) with increasing radius and starts to deviate from the BB relationship 
for $R > 0.3 R_{L1}$. 
The observed trend in outer disc colors is consistent with that inferred from 
the brightness temperature distributions and again suggests an $H$-band excess 
flux for the outer disc regions. The disc colors move toward the upper left of 
the diagram, in the direction opposite to that expected for optically thin gas.
While the error bars are large and the diagram should be viewed with 
caution, the results suggest that the outer disc of IP Peg is opaque with a 
vertical temperature gradient.

\begin{figure}	
\begin{center}
\scalebox{1}[1.00]{%
  \includegraphics[bb=3cm 2cm 20cm 26.cm,scale=0.4]{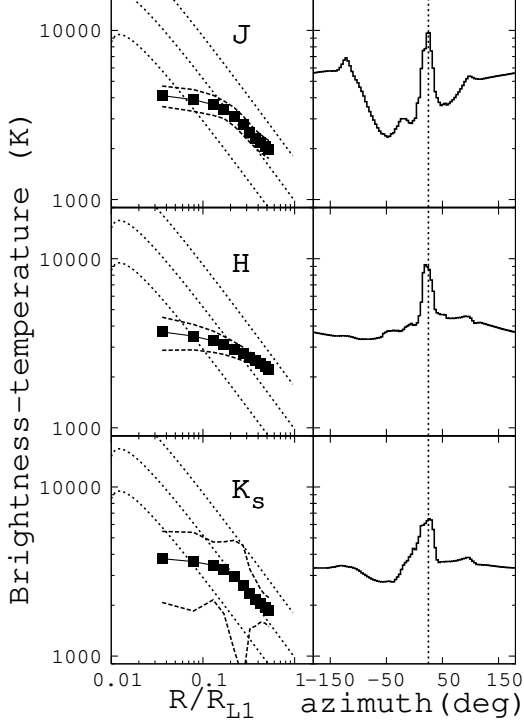} }
\caption{\textbf{Left:} Disc temperature brightness-distribution on a log scale
for the symmetric disc component. The solid curve connecting the filled squares
is the radial profile for the disc and the dashed curves indicate the
1-$\sigma$ limits obtained with Monte Carlo simulations. Dotted lines show 
opaque steady-state disc models for mass accretion rates of 
\.{M}$= 10^{-10}, 10^{-11}$ and $10^{-12}M_{\odot} yr^{-1}$, assuming 
$M_1 = 1.05\,M_{\odot}$ and $R_1 = 0.0124 R_{\odot}$.
\textbf{Right:} Disc rim brightness temperature distribution on a log scale. 
A dotted line indicates the azimuthal position of the bright spot.
}
\label{bri_prof}
\end{center}
\end{figure}

}

\subsection{The secondary eclipse}
\label{sec:ec2}
{

Figs. \ref{cur_luz} and \ref{ec2} shows the model curve from the ellipsoidal 
variation together
with the data. It is clear from the $J$ and $H$-band data that some occulting 
medium is covering part of the light from the secondary star between phases 
0.45 and 0.55. Eclipses of the secondary star by the disc were seen
in the $H$-band data of F99 -- who pointed out that the
shallow depth of the eclipse requires that part of the intervening 
accretion disc must be optically thin -- and also in the spectroscopic data of 
Beekman et al. (2000) -- who deduced that the accretion disc of IP Peg must be
opaque to account for the depth of the secondary eclipse. Littlefair et al. 
(2001) found evidence of an eclipse of the secondary star by optically thin 
parts of the accretion disc from their $K$-band time-resolved spectroscopy of
IP Peg in quiescence.  Assuming local thermal equilibrium, they found that the
occulting optically thin gas is quite hot (10000 K < T < 20000 K) and 
accounts for, at least, the outermost 20 per cent of the disc.  Here we 
model the secondary eclipse to infer the colors of the occulted secondary star
face and to estimate the radial extent of the opaque occulting disc.

Assuming the binary geometry derived in section \ref{sec:ellip}, we simulated 
the eclipse of the secondary star by an opaque cylinder of radius $r_d$ and
half-opening angle $\alpha$ and searched for the values of $r_d$ and $\alpha$
which minimize the $\chi^2$ of the fit to the data.  
In doing this we are making the simplistic assumption that the disc is fully 
opaque up to a certain radius and becomes fully transparent thereafter. If 
opacity changes with radius and height above the disc mid-plane one 
expects 
that a sizeable fraction of the disc will partially transmit the light from 
the secondary star and a larger region would be needed to occult the same 
amount of light. Therefore, our exercise provides a rough lower limit
to the size of the opaque occulting disc.
Having this in mind, the best-fit solutions
are $r_d(J) = 0.19 \pm 0.06\; R_{L1}$, $\alpha(J) = (0.7 \pm 0.4)^{o}$ 
(corresponding to an occultation of $2.3 \pm 0.4$ per cent of the secondary 
star light) and $r_d(H)= 0.22 \pm 0.03\; R_{L1}$, $\alpha(H) = 
(1.5 \pm 0.7)^{o}$ (occultation of $3.0 \pm 0.6$ per cent of the secondary star
light), respectively, for the $J$ and $H$ bands.  The secondary eclipse and 
the best-fit model curves are shown in Fig. \ref{ec2}. The model provides a 
good description of the data for phases $\phi<0.55$ but deviates after the 
eclipse because of the (unaccounted) contribution of the orbital hump to the 
light curve.  Assuming a disc radius of 0.58 $R_{L1}$, this suggests that at 
least the inner $\simeq 1/3$ of the accretion disc is opaque in IP Peg in 
quiescence.

We fitted a blackbody model to the extracted $J$ and $H$ band deficit
fluxes at mid-secondary eclipse to find a temperature of the occulted, 
inner hemisphere of the secondary star of $T_{oc}= 2300 \pm 600\; K$. 
After correcting for the gravity darkening effect, this becomes 
$T_{oc}'= 2500 \pm 700\; K$, consistent at the 1$\sigma$ limit with the 
blackbody temperature of the 
(outer hemisphere of the) secondary star derived in section \ref{sec:ellip}.  
The agreement between the inferred temperatures of the inner and outer 
hemispheres of the secondary star is in line with the conclusion derived from
the ellipsoidal variation fit, namely, that illumination effects on the 
secondary star of IP Peg in quiescence are negligible in the NIR continuum.  
The occulting area
corresponds to $\geq 13$ per cent of the projected surface of the 
secondary star at phase 0.5.
}

}

\section{Discussion}
\label{sec:disc}
{

F99 found a double-humped modulation in their $H$-band light curve 
after subtraction of the ellipsoidal variation from
the secondary star (see their Fig.5). There is no evidence of 
double-humped modulation in our data. Since the phasing of their data
was not secured by a clear identification of the white dwarf egress
feature, their analysis may be affected by errors in the assumed binary phases.
F99 added a phase offset of $\Delta\phi = +0.027$ to their light curves
to correct for deviations of the ephemeris. However, precise measurement 
of white dwarf eclipse egress times from contemporary observations of 
IP Peg in quiescence indicate a phase offset of $\Delta\phi = +0.008$
for the epoch of the F99 observations (Baptista et al. 2005). Thus, the light 
curves of F99 are systematically shifted by $\delta\phi = +0.019$ towards 
positive phases. Indeed, there is an upward kink in the flux at phase $\simeq 
+0.06$ in their data (best seen in their Fig.11) which seemingly
corresponds to the white dwarf egress, suggesting a phase
offset $\simeq +0.015$ with respect to the expected white dwarf 
egress phase.  Simulations show that such a small phase offset between 
the data and the ellipsoidal variation model would be
enough to introduce a spurious double-humped modulation in the light 
curve after removal of the contribution from the secondary star. 
The phase offset also displaces the whole disc brightness distribution 
towards the trailing side of the disc (the one containing the gas 
stream) and would account for much of the offset between the expected 
position of the BS and the position derived from their eclipse map.

\begin{figure}	
\begin{center}
\scalebox{1}[1.00]{%
\includegraphics[bb=1cm 5cm 20cm 24.5cm,scale=0.35,angle=270]{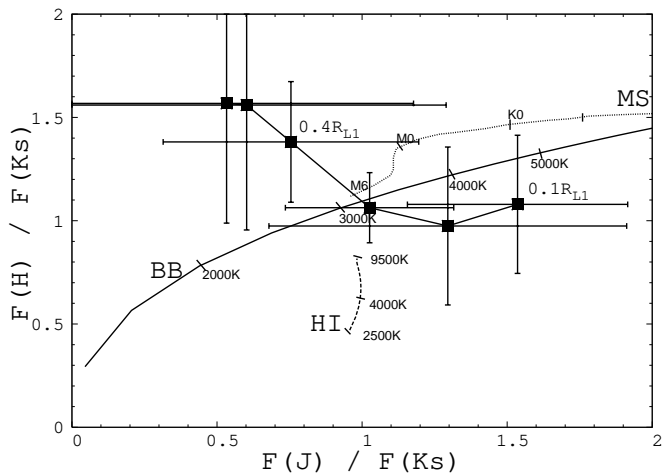} }
\caption{Flux-ratio diagram. A solid line indicate the colors of black body 
emitters. Illustrative temperatures along the sequence are indicated by labels.
A dashed line shows the colors of optically thin emitting gas with three  
temperatures labeled. A dotted line indicates stellar main-sequence 
colors, with some spectral types labeled. Filled squares connected by solid 
lines indicate the color of the symmetric disc component. The error bars 
represent the standard deviation with respect to the mean value in each radial 
bin. Labels mark the radius in units of $R_{L1}$.}
\label{ccd}
\end{center}
\end{figure}

The modeled ellipsoidal variation also indicates that the irradiation of
the secondary star surface is negligible in the infrared.
This result is in apparent contradiction with those of Davey \& Smith (1992)
and Watson et al (2003). Despite
the differences in the results, both works find a significant decrease 
in NaI $\lambda 8190$\AA ~line strength over the inner face of the secondary 
star in IP Peg as a consequence of irradiation effects. We remark that, if the 
irradiated energy does not penetrate deeply in the atmosphere but mainly heats
the upper atmospheric layers, it may reduce the
vertical temperature gradient (and, therefore, lead to a decrease in NaI
absorption line strength) without affecting the continuum radiation arising 
from deeper atmospheric layers. This is in line with the investigation of 
irradiation effects on CV secondaries by Barman, Hauschildt \& Allard (2002).
Their preliminary study shows that, for a typical CV, irradiation leads to a 
significant change in the temperature structure of the uppermost atmospheric 
layers of the secondary star (where the NaI line comes from) leaving the deeper
layers (the $\tau = 1$ region where the IR continuum is produced) mostly 
unaffected.

\begin{figure}	
\begin{center}
\scalebox{1}[1.00]{%
\includegraphics[bb=2cm 3.5cm 20cm 26.5cm,scale=0.5]{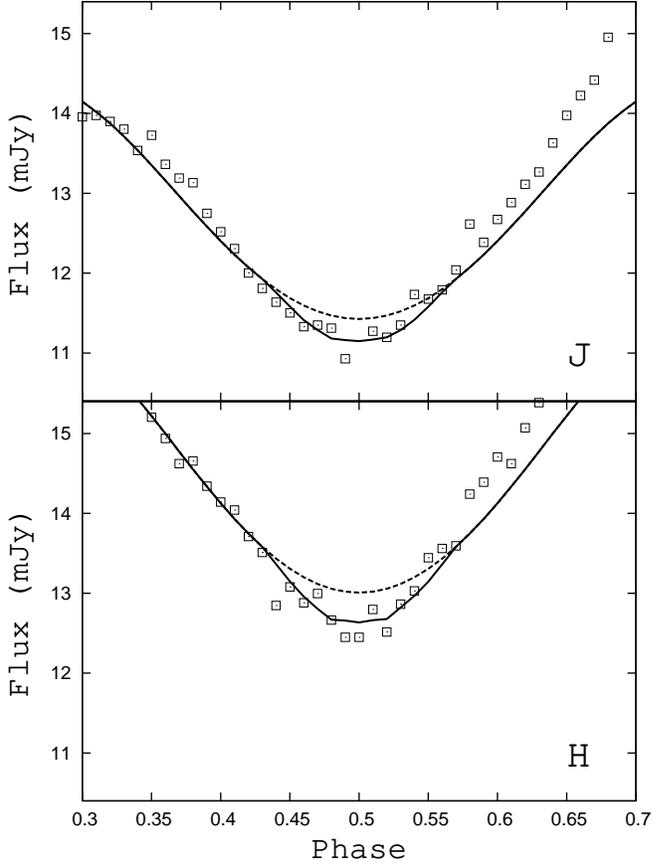} }
\caption{$J$ and $H$-band light curves of IP Peg (open symbols) with modeled 
ellipsoidal variation without (dashed line) and with (solid lines) secondary 
eclipse included.}
\label{ec2}
\end{center}
\end{figure}

Our disc temperatures are higher than those inferred by F99. We find 
temperatures of $\sim3500K$ in all bands while they find temperatures of 
$\sim 3000K$ in $H$. This is in line with the fact
that our $H$-band light curve has 2.1 mJy more flux than the 1994 Sep
light curve of F99 (at the reference phase $\phi= 0.25$, where the
contribution of BS anisotropic emission should be minimal). Our data
and those of F99 were collected, respectively, 5 and 6 weeks after an
outburst. The observed difference in disc temperatures could be
explained if the accretion disc of IP Peg slowly cools
down during quiescence. Alternatively, the observed difference in light curve 
flux and disc temperatures could be accounted for by the $10\%$ uncertainty in 
flux calibration of both works.
}

\section{Summary}
\label{sec:concl}
{

We modeled the ellipsoidal variations in $JHK_{s}$ light curves of IP Peg to 
estimate the system parameters and the contribution of the secondary to the
total light (Table \ref{tab_result}). Using the calibrated fluxes we inferred
an equivalent blackbody temperature of $3100\pm500\;K$, and we derived a distance 
to the system of $115\pm30$ pc. 

Using 3D eclipse mapping techniques we derived the surface brightness 
distribution of the IP Peg accretion disc. With the aid of the entropy of the 
eclipse map, we find a disc half-opening angle of $\alpha = 2^{o}$. The disc 
brightness distribution shows an asymmetric structure along the gas stream 
trajectory in the inner disc regions close to the white dwarf in all bands 
($R \sim 0.1 R_{L1}$), indicating the existence of gas stream overflow in 
IP Peg.

The amplitude of the orbital hump decreases with increasing wavelength. The hump is 
modeled as an extended ($\Delta \theta \simeq 30^o$) bright spot located at the
edge of the accretion disc with brightness temperatures of $\sim 10000K$ in
$J$ and $H$ and $\sim 6000K$ in the $K_{s}$ band. The phase of maximum of the 
orbital 
hump in NIR-light curves, $\phi = 0.74$, does not coincide with the one in the 
optical ($\phi = 0.9$, see Fig. 1 of Wood \& Crawford 1986) nor does it 
coincide
with the phase of maximum emission of a radially emitting bright spot. The 
eclipse position of the bright spot differs from its theoretical azimuthal 
position by $\sim 5^o$ and its direction of maximum emission (inferred from the
light curve) differs by $\sim 54^o$ from the direction of radial emission. 

In order to match the depth and width of the secondary eclipse, 
a sizeable fraction of the accretion disc must be opaque. 
The inferred temperatures of the inner and outer hemispheres
of the secondary star are the same to within the uncertainties, 
indicating that illumination effects are negligible for the IR continuum.

}

\section{Acknowledgments}
{
We thank Don Hoard for pointing out the reference for the MS IR colors, Frank 
Gribbin and Chris Benn for useful information about the WHIRCAM, and the 
anonymous referee for useful comments which helped to improve the 
presentation of our results.
This work was partially supported by CNPq (Brazil) through research grant
62.0053/01-1-PADCT III/Milenio. T.R. acknowledges financial support from
CNPq. R.B. acknowledges financial support from CNPq (Brazil)
through grants 300.354/96-7 and 200.942/2005-0.
This publication makes use of data products from the Two Micron All Sky 
Survey, which is a joint project of the University of Massachusetts and the 
Infrared Processing and Analysis Center/California Institute of Technology, 
funded by the National Aeronautics and Space Administration and the National 
Science Foundation.
In this research we have used, and acknowledge with thanks,
data from the AAVSO International Data base, which are based
on observations collected by variable star observers worldwide.
}


\begin{thebibliography}{9}

\bibitem {bap1} Baptista, R. \& Steiner, J. E. 1993, {A\&A}, 277, 331.
\bibitem {bap05} Baptista, R., Morales-Rueda L., Harlaftis, E.T., Marsh, T.R., \&
Steeghs, D. 2005, A\&A, 444, 201.
\bibitem {bha} Barman, T., Hauschildt, P. \& Allard, F., 2002, in The Physics of Cataclysmic 
Variables and Related Objects, ASP Conference Series, Vol. 261, ed.
B. T. Gaensicke, K. Beuermann, \& K. Reinsch. p. 49. 
\bibitem {beek} Beekman, G., Somers, M., Naylor, T. \& Hellier C. 2000, 
  {MNRAS}, 318, 913.
\bibitem {bb88} Bessell, M. S. \& Brett, J. M. 1988, PASP, 100, 1134. 
\bibitem {bbc_07} Borges, B.W., Baptista, R.\& Catalán, M.S., 2007, in preparation.
\bibitem {c01} Carpenter, J. M. 2001, AJ, 121, 2851.
\bibitem {clr} Claret, A. 1998, {A\&A}, 335, 647.
\bibitem {ds92} Davey, S. \& Smith, R.C. 1992, {MNRAS}, 257, 476
\bibitem {diaz} Diaz-Cordovés, J. \& Giménez, A. 1992, {A\&A}, 259, 227.
\bibitem {fro} Froning, C.S., Robinson, E.L., Welsh W.F. \& Wood, J.H., 1999, {AJ},
  523, 399 (F99).
\bibitem {ha_etal05a} Harrison, T. E.; Osborne, H. L.; Howell, S. B. 2005a, AJ, 129, 2400.
\bibitem {ha_etal05b} Harrison, T. E.; Howell, S. B.; Johnson, J. J. 2005b, 
Bull. Am. Astron. Soc., 207, 7016.
\bibitem {hrd96} Hughes, S. M., Roche, P., Dhillon, V. S., 1996, WHIRCAM User's
      Guide v1.0, Isaac Newton Group of Telescopes, La Palma
\bibitem {hg00} Huré, J.-M., Galliano F. 2001, A\&A, 366, 359.
\bibitem {k79} Kurucz, R. L. 1979, {ApJS}, 40, 1.
\bibitem {leg} Leggett, S.K. 1992, {ApJS}, 82, 351.
\bibitem {lit} Littlefair, S.P., Dhillon, V.S., Marsh, T.R. \& Harlaftis E. T., 2001,
  {MNRAS}, 327, 475.
\bibitem {mh90} Marsh T. R. 1988, MNRAS, 231, 1117.
\bibitem {mjs87} Martin, J.S., Jones, D.H.P \& Smith, R.C. 1987, {MNRAS}, 224, 1031.
\bibitem {mmh82} Meyer, F. \& Meyer-Hofmeister, E. 1982, A\&A, 106, 34.
\bibitem {n72} Nauenberg, M. 1972, ApJ, 175, 417.
\bibitem {Press} {Press, W.H., Flannery, B.P., Teukolsky, S.A., Vetterling, 
W.T., 1986,}{ Numerical Recipes, Cambridge University Press}
\bibitem {rutt_etlal92} Rutten R.G.M., van Paradijs J. \& Tinbergen J., 1992,
A\&A, 254, 159.
\bibitem {sb06} Saito, R. K. \& Baptista, R. 2006, AJ, 131, 2185.
\bibitem {S89} Sarna, M. J. 1989, A\&A, 224, 98.
\bibitem {s92} Smak, J. I. 1992, IAUS, 151, 83.
\bibitem {sr98} Spruit, H. C. \& Rutten R.G.M. 1998, MNRAS, 299, 768.
\bibitem {setal79} Strecker, D.W., Erickson, E.F. \& Whitteborn, F.C., 1979,
ApJS, 41, 501.
\bibitem {s00} Skidmore, W., Mason E., Howell, S. B., Ciardi, D. R., Littlefair S. P., Dhillon V. S., 2000, MNRAS, 318, 429.
\bibitem {sk_etal06} Skrutskie M.F. et al. 2006, AJ, 131, 1163.
\bibitem {szk} Szkody, P. \& Mateo M. 1986, {AJ}, 92, 483.  
\bibitem {war} Warner, B. 1995, {Cataclysmic Variables Stars.},
  Cambridge Univ. Press, Cambridge.
\bibitem {wdrs03} Watson, C. A., Dhillon, V. S., Rutten, R. G. M., Schwope, A. 
D., 2003, MNRAS, 341, 129.
\bibitem {wolf} Wolf, S., Mantel, K. H.;Horne, K.; Barwig, H. Shoembs, 
  R.; Baernbantner, O., 1993, {A\&A}, 273, 160.
\bibitem {wc96} Wood, J. H. \& Crawford, C. S. 1986, {MNRAS}, 222, 645.
\bibitem {wetal86} Wood, J. H., Horne K., Berriman G., Wade R. A., O'Donoghue D., Warner B., 1986, MNRAS, 219, 629. 
\bibitem {wetal89} Wood, J. H., Horne K., Berriman G., Wade R. A., 1989, ApJ, 
341, 974.

\end{thebibliography}
\end{document}